\documentclass{article}
\usepackage{LaThuileFPSpro}
\usepackage[dvips]{graphicx}
\usepackage{amssymb}
\begin{document}
\title{ 
  SEARCH FOR LOW MASS SM HIGGS AT THE TEVATRON
  }
\author{
  Michiel P. Sanders \\
  {\em LPNHE/IN2P3/CNRS, Paris, France} \\
  For the D\O\ and CDF Collaborations        \\
  }
\maketitle

\baselineskip=11.6pt

\begin{abstract}
  The only place in the world where at this time standard model Higgs
  bosons can be produced and detected is the Tevatron at Fermilab. In
  this contribution, the most recent results on the search for a low
  mass Higgs boson are presented, using datasets of up to
  $1.9\,\mathrm{fb}^{-1}$. In the absence of signal, the
  combined Tevatron cross section limit at a Higgs boson mass of 115
  GeV is determined to be 6.2 (4.3 expected) times the standard model
  (SM) expectation, at 95\%
  confidence level. The expected gain in sensitivity from the
  forthcoming larger dataset and improved analysis methods will likely
  make an exclusion or
  observation at low mass possible in the near future.
\end{abstract}
\newpage
\section{Introduction}
The standard model of particle physics as we know it has been very
successful. Many precision measurements have given excellent agreement
with the model, and many processes predicted by the standard model
have been observed. However, the success of the standard model depends
on a mechanism to break the electroweak symmetry. Without that, the W
and Z bosons would remain massless.

The Higgs mechanism is the most promising way to break the electroweak
symmetry. It gives mass to the electroweak bosons and it leaves the
photon massless. The same Higgs field can be used to give mass to the
quarks and leptons. An essential prediction of the Higgs mechanism is
the existence of a yet unobserved particle: the Higgs boson (H). 

Through radiative corrections, the mass of the top quark and the mass
of the W boson depend on the mass of the Higgs boson. Precision
measurements of these
parameters, and many others, at LEP, SLD and the Tevatron can thus be
interpreted in the standard 
model as a prediction for the mass of the Higgs boson. At the time of
this meeting, the central value for this prediction\cite{EWfit} was
$m_{\mathrm{H}} = 76^{+33}_{-24}$ GeV, leading to an upper limit 
\footnote{All limits quoted in this contribution are
given at 95\% C.L.} of 144 GeV. 
Including the direct Higgs mass limit from LEP\cite{LEPhiggs}
of 114.4 GeV raises this upper limit to 182 GeV. The Higgs boson is
thus expected to have a relatively low mass, within reach of the
Tevatron experiments. 

\section{Low mass Higgs and the Tevatron}

The Tevatron is a $p\bar{p}$ collider at Fermilab, near Chicago,
running at a centre-of-mass energy of 1.96 TeV. The two
general-purpose experiments D0 and CDF have collected a data sample
corresponding to an integrated luminosity of about
$3.7\,\mathrm{fb}^{-1}$. The results shown in the following are based on
datasets of up to $1.9\,\mathrm{fb}^{-1}$. 

At the Tevatron, the dominant production mode for a low mass Higgs
boson ($m_{\mathrm{H}} \lesssim 140$ GeV)
is the gluon fusion process. The dominant decay mode for the
Higgs boson  
is to the kinematically allowed heaviest particle, in this case
to a $\mathrm{b}\bar{\mathrm{b}}$ quark pair. Experimentally the
combination of gluon fusion and 
$\mathrm{b}\bar{\mathrm{b}}$ decay is unfeasible due to the enormous
background from dijet production.
The next most dominant production modes are those where the Higgs is
produced in association with a W or Z boson. In this case the
$\mathrm{H}\to\mathrm{b}\bar{\mathrm{b}}$ decay mode is accessible,
using the leptonic or invisible decay modes of the W and Z bosons. 

The search for a low mass Higgs boson at the Tevatron is thus a search
for a pair of jets originating from b-quarks in association with a
leptonic or invisible W
or Z signature. 
For $\mathrm{WH}\to\ell\nu \mathrm{b}\bar{\mathrm{b}}$, the expected
cross section times branching ratio (for one lepton flavour) is of
the order of 20 fb at a Higgs mass of 105 GeV down to 4 fb at
140 GeV. The same quantity for ZH production with the Z decaying
invisibly to neutrinos is essentially the same as that for WH for one
lepton flavour. ZH production with the Z decaying to an e, $\mu$, or
$\tau$ pair is about a factor of five below that.

Tagging a jet as originating from a b-quark is an important ingredient
of this search. Both experiments use combinations of 
variables sensitive to the presence of B-mesons, such as a reconstructed 
secondary vertex, large track impact parameters
with respect to the primary vertex and the secondary vertex mass, to
obtain efficient b-tagging algorithms. As an example, D0's neural net
tagger\cite{TScanlon} obtains a 50\% b-tag efficiency for a mis-tag rate of 0.5\%
(``tight'' tag) and 74\% efficiency at 5\% mis-tag rate (``loose''
tag). Typically, in analyses either two loose b-tags or one tight
b-tag are required (exclusively). 

\section{Low mass Higgs searches at the Tevatron}

In the following sections, the current state of the various low
mass Higgs searches at D0
and CDF is described.
Common to all analyses is the clean signature of the decay of a Z or W
boson as one or two high momentum leptons (only electrons and muons are
considered; leptonic $\tau$ decays are typically included) and/or
large missing transverse energy due to undetected neutrinos. The Higgs
boson decay has the signature of two or more (due to initial or final
state radiation) jets. 

Background sources include production of a W or Z boson in
association with jets (including b-quark jets), production of
top-quark pairs or a single top, and diboson production (WW, WZ,
ZZ). The amount of background from these processes is estimated from Monte Carlo simulation
of the physics process (with generators like \textsc{alpgen},
\textsc{pythia} and \textsc{herwig}) and the detector
response. Another source of background is that where a jet in a multijet event,
which are abundantly produced, is misidentified as a lepton from a W
or Z boson decay, and where energy mismeasurements lead to
missing transverse energy. This source of background is typically
estimated from the data itself. 
 
\subsection{Two charged leptons: $\mathrm{ZH}\to\ell\ell \mathrm{b}\bar{\mathrm{b}}$}

The event signature of $\mathrm{ZH}\to\ell\ell
\mathrm{b}\bar{\mathrm{b}}$ production is very clean: two isolated
leptons with an invariant mass close to the Z boson mass, and two
jets. Both D0 and CDF have analyzed datasets corresponding to $\simeq
1 \,\mathrm{fb}^{-1}$ of integrated luminosity. No updates were made
recently. 

The CDF collaboration applies a constrained fit using the measured jet
energies and missing transverse energy to improve the jet energy
measurement. To gain acceptance, D0 has rather low transverse momentum
cuts on the leptons. In spite of these differences between the
analyses, the final sensitivity to a Higgs boson signal is similar. 
Both collaborations use neural networks to improve the sensitivity
over that obtained using the dijet invariant mass. 

\begin{figure}[htb]
\hfill
\includegraphics[height=.35\textwidth,width=.45\textwidth]{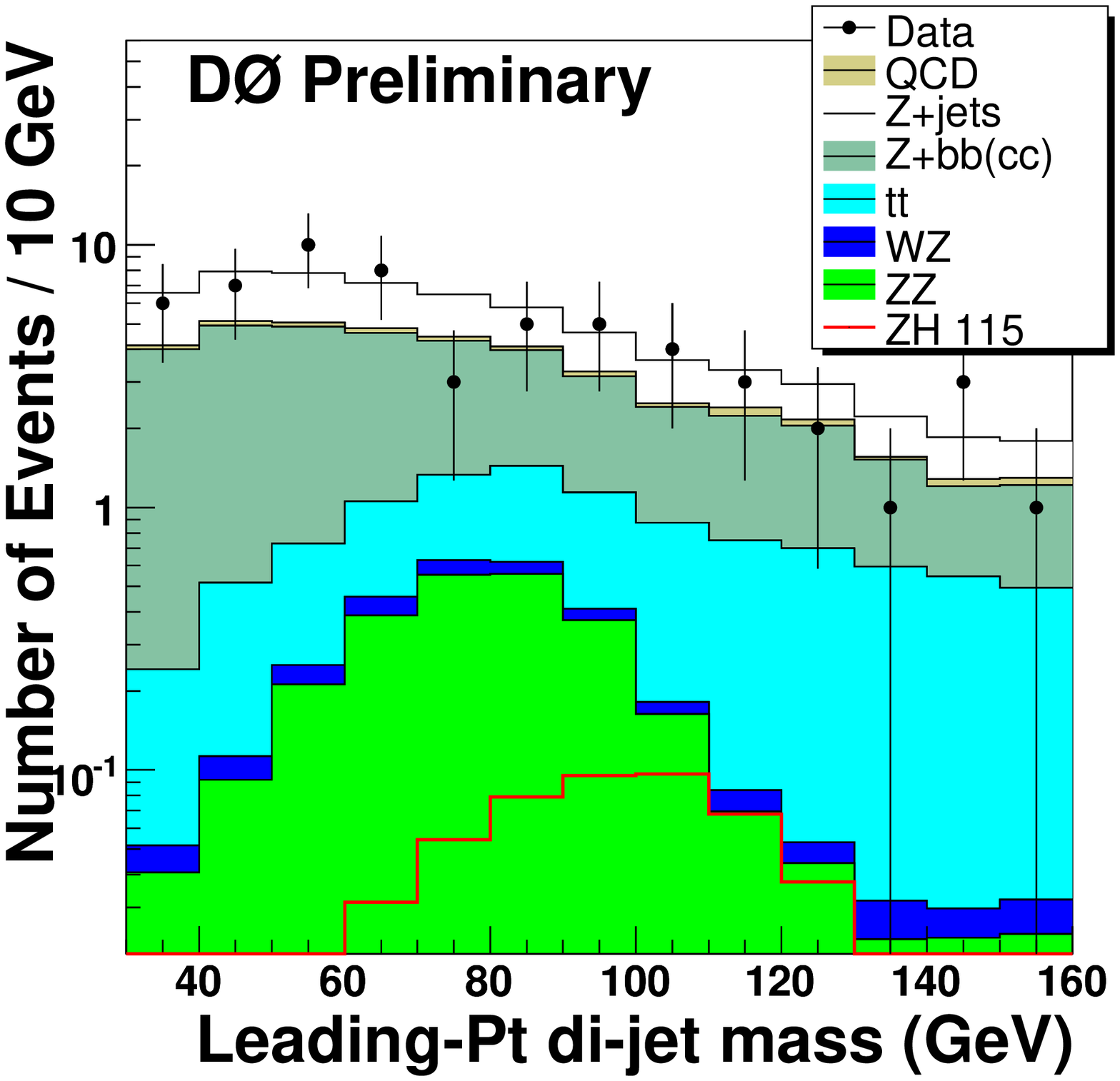}
\hfill
\includegraphics[height=.35\textwidth,width=.45\textwidth]{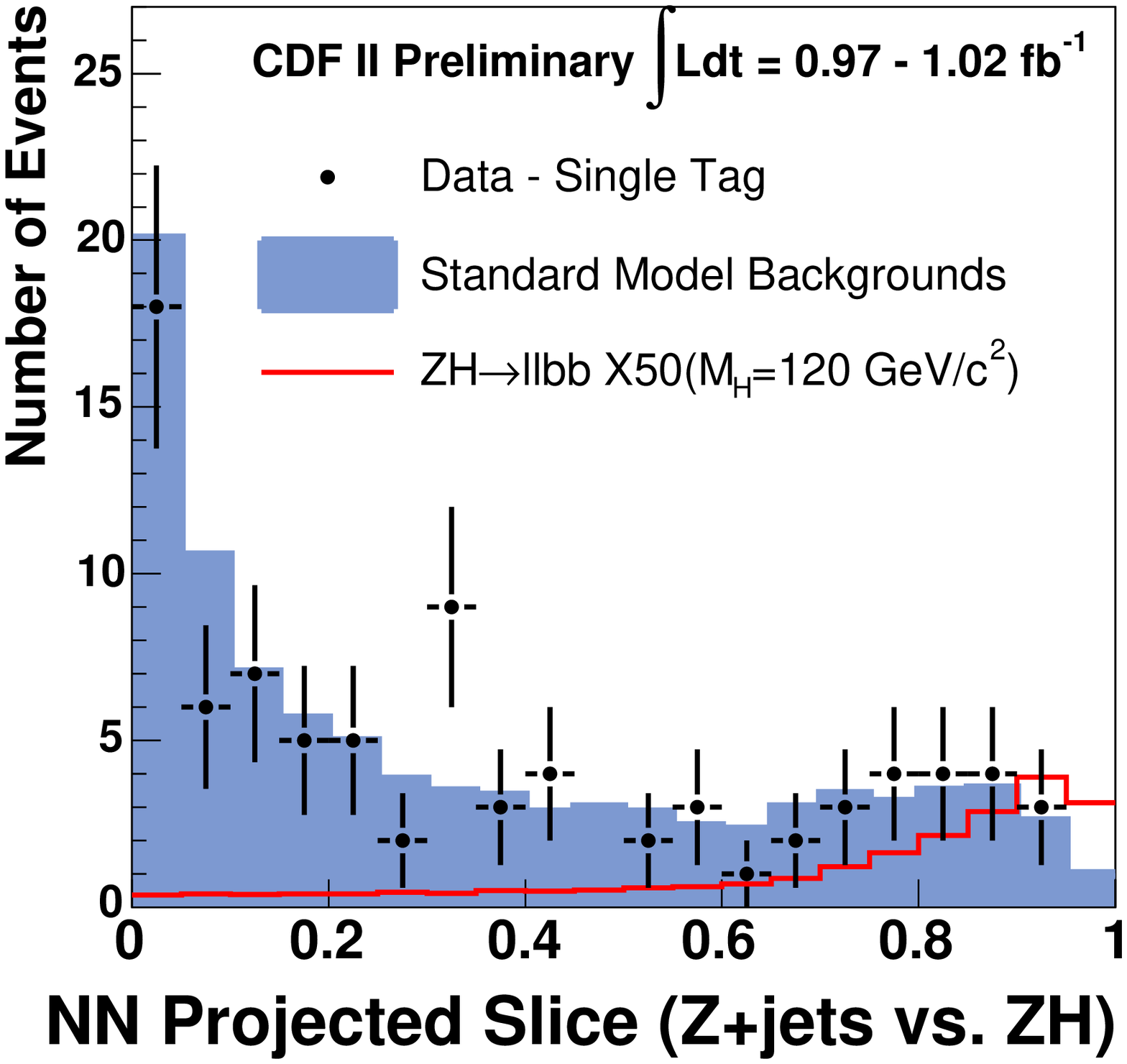}
\hfill
\caption{Dijet mass distribution obtained by D0 in the
  $\mathrm{ZH}\to\ell\ell\mathrm{b}\bar{\mathrm{b}}$ analysis after requiring two
  b-tags in the event (left) and a slice of CDF's neural network
  output in
  the same search channel (right).
\label{fig:ZH}}
\end{figure}

The dijet mass distribution found by D0, after requiring both jets to
be b-tagged is shown in fig.\ref{fig:ZH} (left). The expected Higgs boson mass
peak is clearly visible, but the amount of background, in particular
from Z$+\mathrm{b}\bar{\mathrm{b}}$ production is large. CDF uses a
two dimensional neural network, trained against Z+jets and top-quark
pair production. A slice of the final CDF
neural network output distribution is shown in
fig.\ref{fig:ZH} (right). The Higgs boson signal clearly peaks towards
large output values. 

Neither D0 nor CDF finds an excess, and the data agree well with the
background model. The neural network distributions are then used to
derive limits on the Higgs boson production cross section. At a
Higgs mass of 115 GeV, D0 finds a limit of 18 (20 expected) times the
expected standard model cross section times branching ratio of $\mathrm{H}\to
\mathrm{b}\bar{\mathrm{b}}$, whereas CDF finds 16 (16 expected).

\subsection{One charged lepton: $\mathrm{WH}\to\ell\nu \mathrm{b}\bar{\mathrm{b}}$}

Removing a lepton from the final state described in the previous
section, and instead requiring some missing transverse energy leads to
the final state corresponding to $\mathrm{WH}\to\ell\nu
\mathrm{b}\bar{\mathrm{b}}$ decays. 

\begin{figure}[htb]
\hfill
\includegraphics[height=.35\textwidth,width=.45\textwidth]{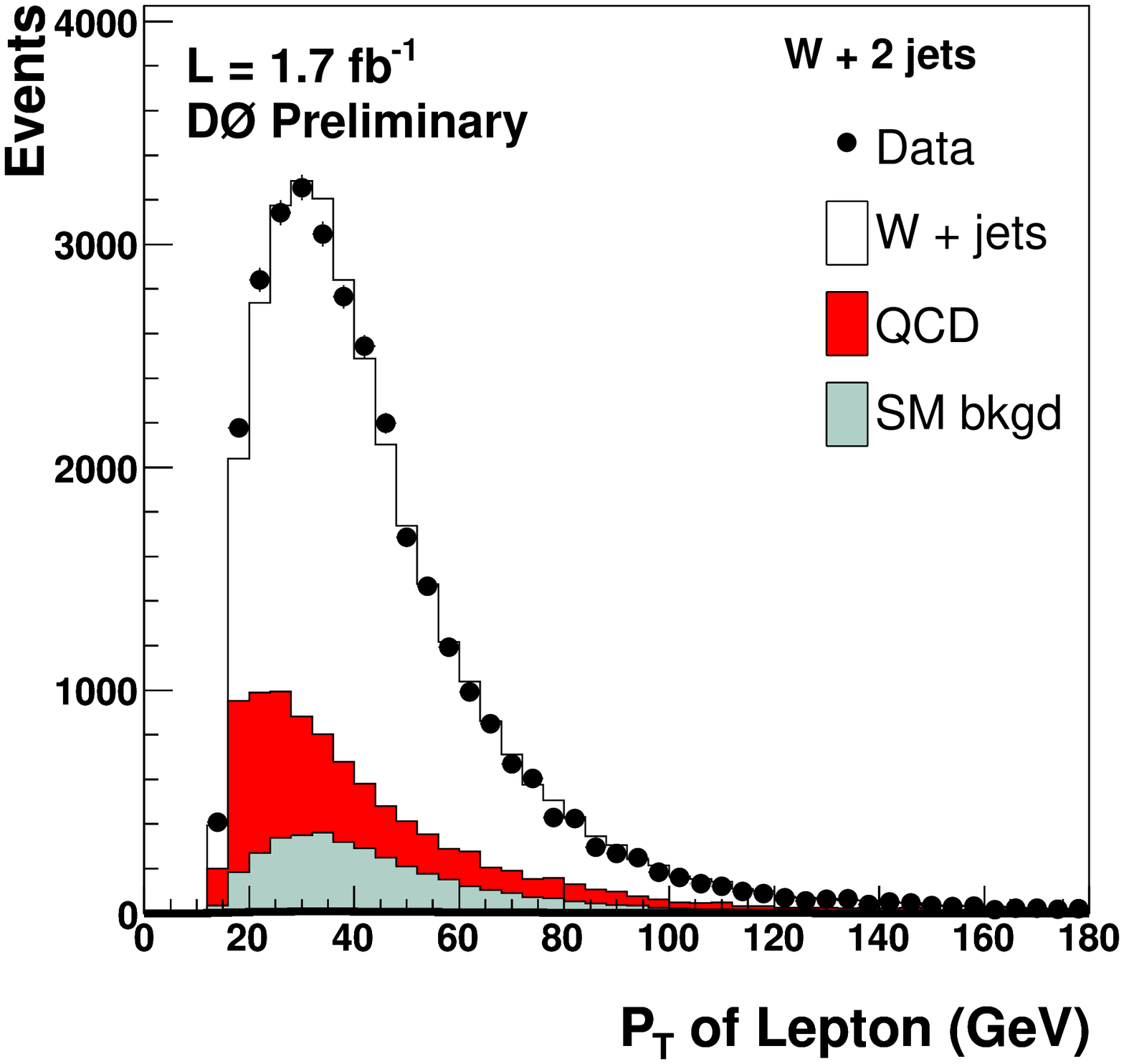}
\hfill
\includegraphics[height=.35\textwidth,width=.45\textwidth]{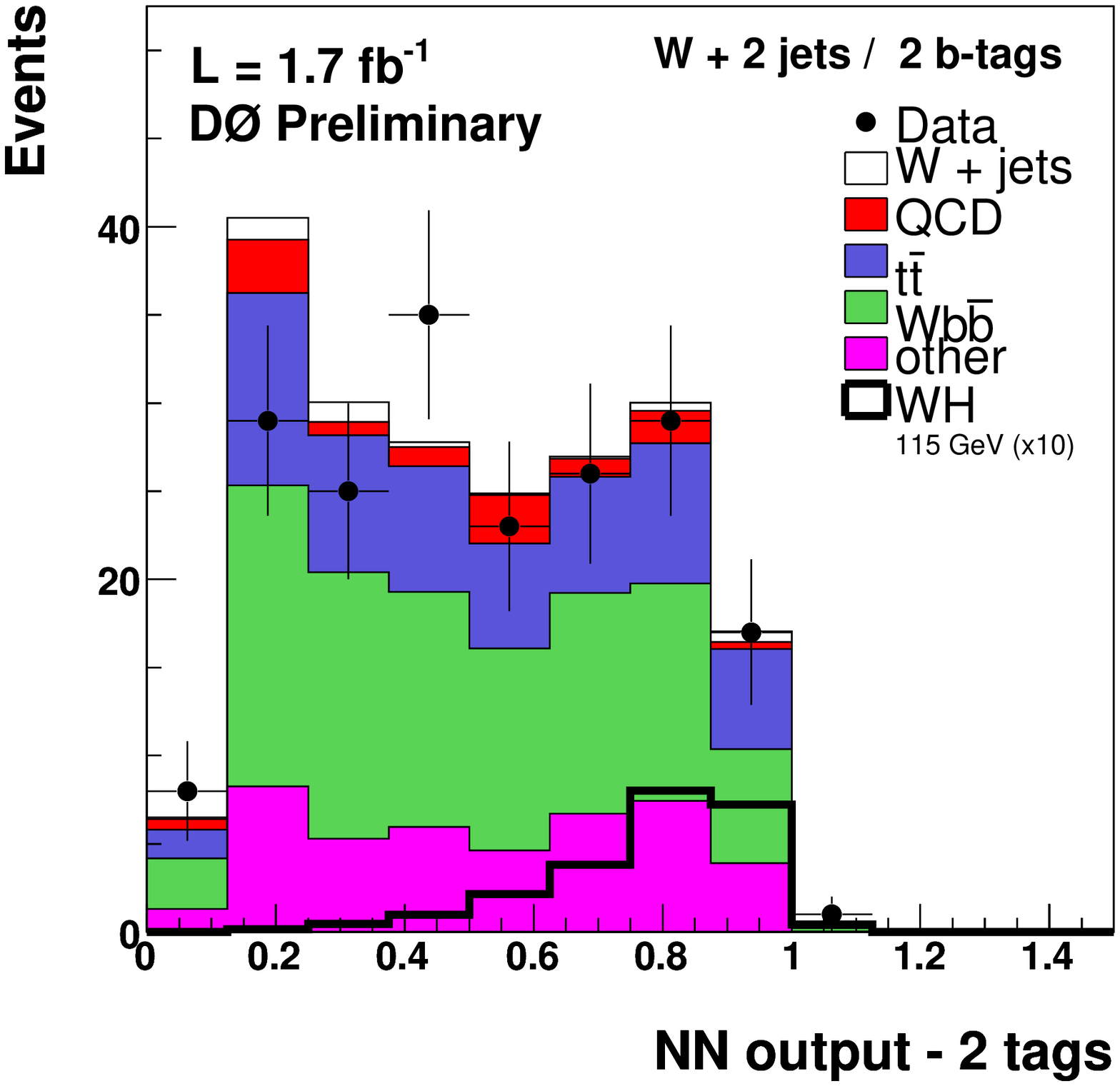}
\hfill
\caption{Distribution of the transverse momentum of the lepton in the
  D0 $\mathrm{WH}\to\ell\nu \mathrm{b}\bar{\mathrm{b}}$ analysis,
  before b-tagging requirements are applied to the two jets (left),
  and the final neural network output distribution for events where
  both jets are b-tagged (right).
\label{fig:WHDZero}}
\end{figure}

In this channel, D0 has analyzed
a data set corresponding to an integrated luminosity of
$1.7\,\mathrm{fb}^{-1}$. Fig.\ref{fig:WHDZero} (left) shows an example
of the total background levels and the contribution from multijet
production, in events with a W boson candidate and two jets. For an
expected WH signal contribution ($m_{\mathrm{H}} = 115$ GeV) of 9.9
events, a background of 33.5k events is expected. The dominant
background source is production of a W boson in association with jets
(white histogram), and the next most significant background source is
that of multijet production (red histogram). The final discriminant
variable in the D0 analysis is the output of a neural network, trained
to separate Higgs boson signal from background. The distribution is
shown in fig.\ref{fig:WHDZero} (right), for events where both jets are
b-tagged. At this analysis stage, the expected number of signal events
is 2.3, with a background of 204 events. 

\begin{figure}[htb]
\hfill
\includegraphics[width=.45\textwidth,height=.35\textwidth]{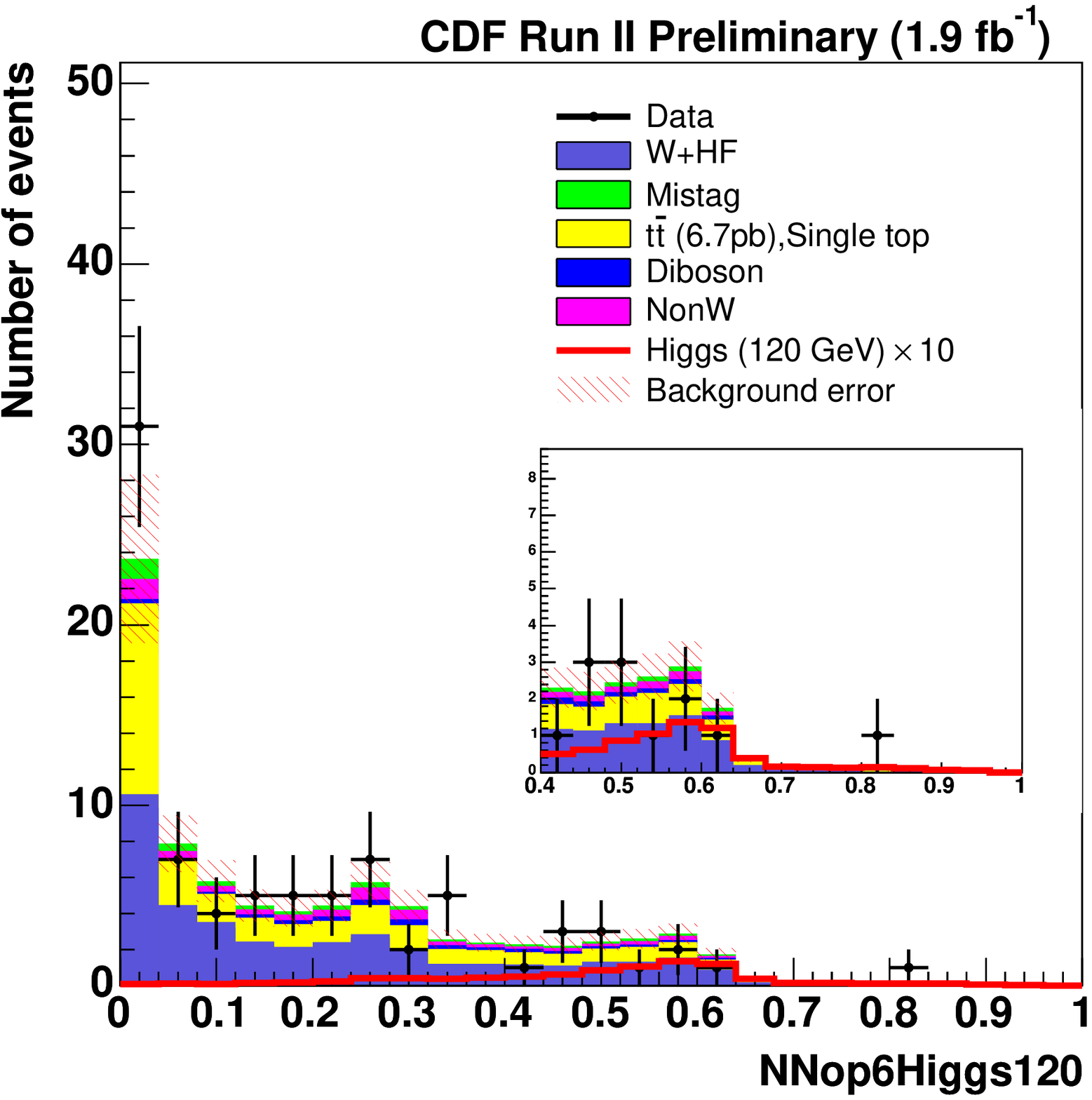}
\hfill
\includegraphics[width=.45\textwidth,height=.35\textwidth]{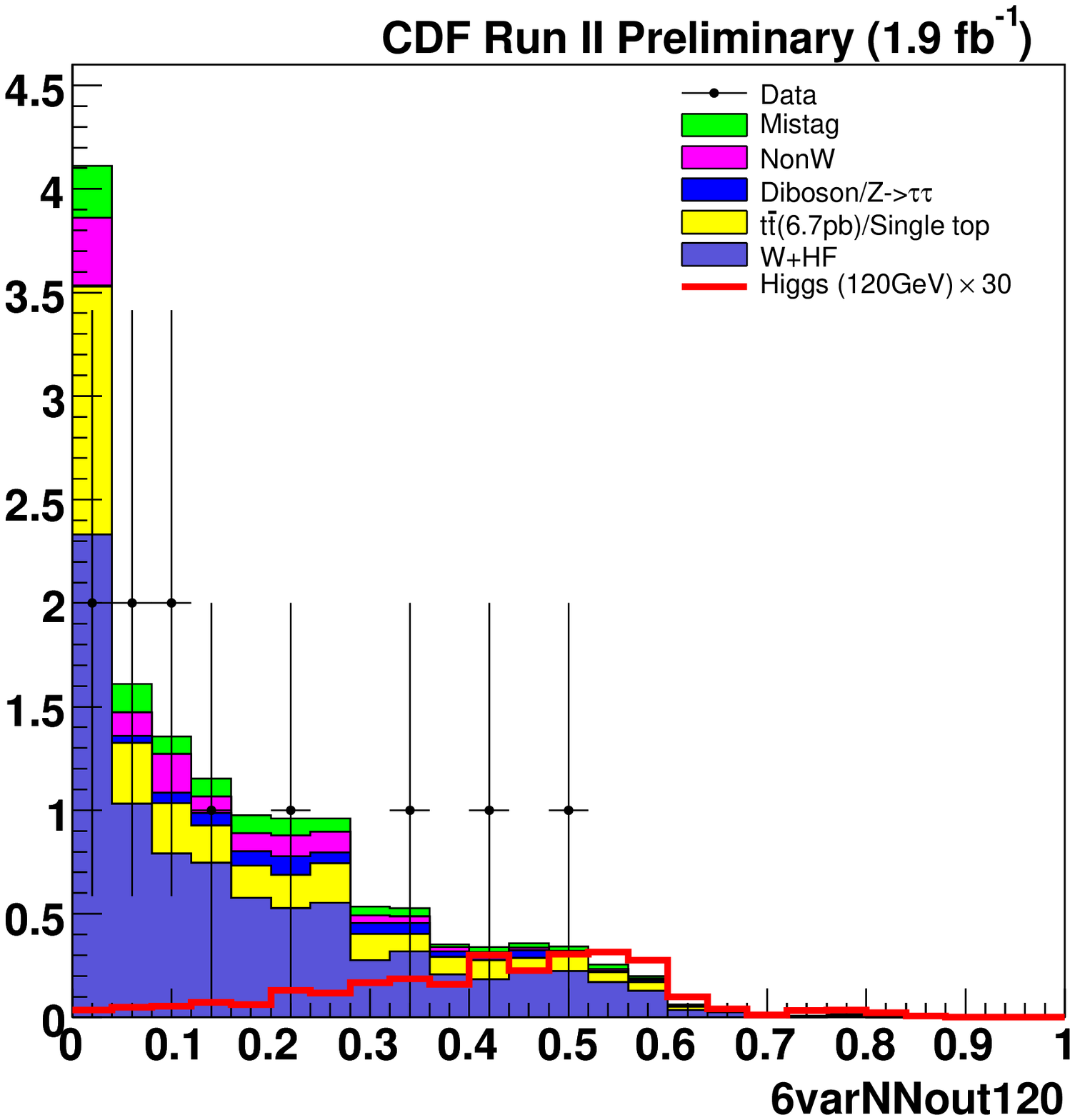}
\hfill
\caption{Neural network output distributions obtained by CDF in the 
$\mathrm{WH}\to\ell\nu \mathrm{b}\bar{\mathrm{b}}$ analysis for
  central electrons and muons (left) and forward going electrons (right).
\label{fig:WHCDF}}
\end{figure}

The CDF collaboration has released new results in this search channel
with a slightly larger integrated luminosity ($1.9\,
\mathrm{fb}^{-1}$). The b-tagging
classification was extended to two double-tag and one single tag category (all
exclusive). Moreover, CDF has already included the forward-going
electrons (``plug'' electrons) in the analysis. An example of that is
shown in fig.\ref{fig:WHCDF} where on the left the final neural
network output distribution is shown for the events with a ``central''
electron or muon, and on the right that for events with a ``plug''
electron. The expectation is to find 0.09 WH events in the plug region
(electron only), to be compared to 0.94 in the central region
(electron and muon combined). This 10\% increase in signal acceptance
comes at the cost of larger relative background levels (14.2 events in
the plug region versus 80.4 in the central region).

Again, both D0 and CDF find good agreement between the data and the
expected background, without any sign of a Higgs boson
signal. Therefore, cross section limits for Higgs boson production are
derived using the final neural network output distributions. D0 finds,
at an assumed Higgs mass of 115 GeV, a limit of 11.1 (9.1 expected)
times the standard model expectation, and CDF find 8.2 (7.3
expected). The CDF result is better than the current D0 result, but an
improved analysis from D0, using forward electrons, three-jet events
and an improved neural network, will be finalized in the near future.

\subsection{No charged leptons: $\mathrm{ZH}\to\nu\nu \mathrm{b}\bar{\mathrm{b}}$}

Removing the lepton from the final state described in the previous
section leads to the signature of $\mathrm{ZH}\to\nu\nu
\mathrm{b}\bar{\mathrm{b}}$ events. 

\begin{figure}[htb]
\hfill
\includegraphics[width=.45\textwidth,height=.35\textwidth]{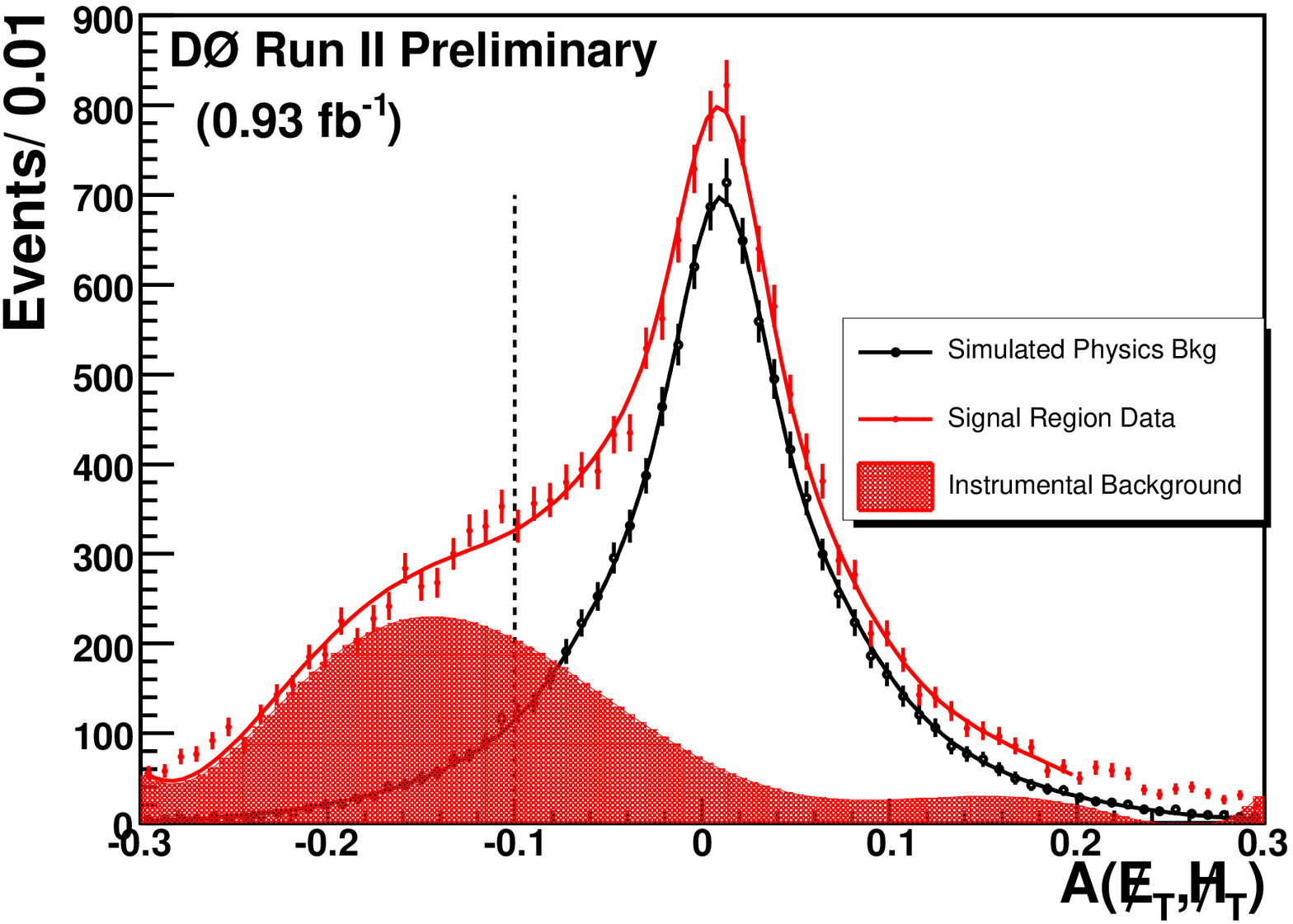}
\hfill
\includegraphics[width=.45\textwidth,height=.35\textwidth]{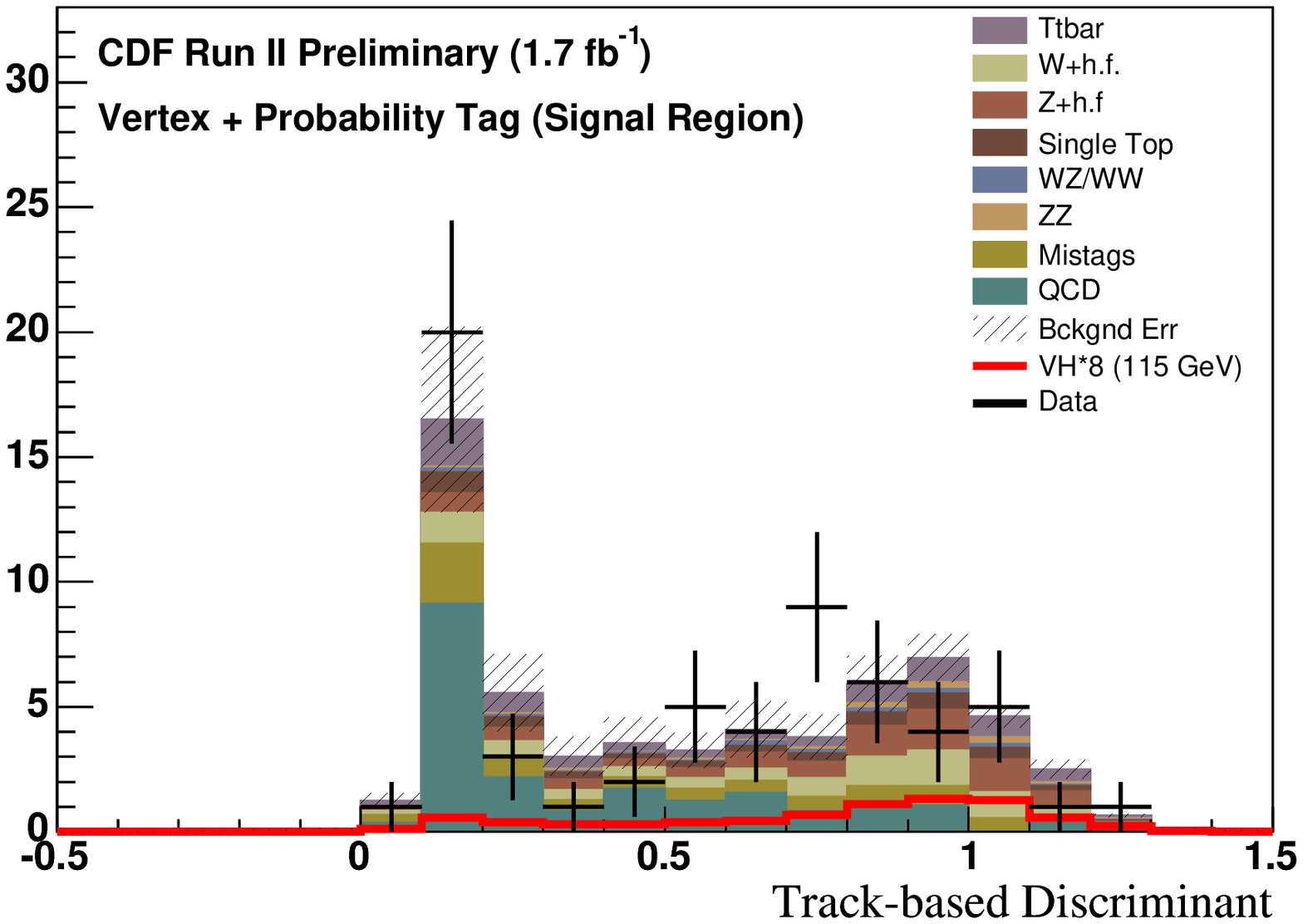}
\hfill
\caption{Distribution of the missing transverse energy asymmetry as
  defined by D0 in the $\mathrm{ZH}\to\nu\nu
  \mathrm{b}\bar{\mathrm{b}}$ analysis, before b-tagging (left) and
  distribution of the neural network output used by CDF to isolate multijet
  production, after b-tagging (right).
\label{fig:ZHnunubb}}
\end{figure}

In this case, the multijet
background in the analysis is caused by events in which the energy of
the jets is mismeasured, leading to missing transverse
energy. The level of this background contribution can be estimated by
using the fact that if the energy of one jet is mismeasured, the missing
transverse energy will point in the jet direction. Also, 
the missing energy calculated using tracks will be different from the
calorimeter based missing energy in the case of a calorimeter
mismeasurement.
The D0 collaboration also uses the asymmetry between missing energy
calculated with all calorimeter information and missing energy using
the reconstructed jets as a measure of multijet production
(as shown in fig.\ref{fig:ZHnunubb} (left)). CDF has used a dedicated
neural network to separate the multijet background from others (shown
in fig.\ref{fig:ZHnunubb} (right)). 

Both D0 and CDF train a neural network to improve the sensitivity of
the analysis, as compared to considering the dijet mass distribution
only. In the absence of a signal excess, D0 finds a cross section limit at a
Higgs boson mass of 115 GeV of 13 (12 expected) times the standard
model expectation, using a relatively small dataset of
$0.9\,\mathrm{fb}^{-1}$.
A new analysis from the CDF collaboration, using an integrated
luminosity of $1.7\,\mathrm{fb}^{-1}$, leads to a limit of 8.0 (8.3
expected) times the standard model expectation. 

\subsection{Tevatron combination}

All analyses presented in the preceding sections specifically search
for a standard model Higgs boson (at low mass). For optimal
sensitivity, the
individual search results of the two experiments can then be combined,
assuming standard model 
branching ratios and cross sections. At the time of this meeting, the
new CDF results in the 
$\mathrm{WH}\to\ell\nu \mathrm{b}\bar{\mathrm{b}}$ and $\mathrm{ZH}\to\nu\nu \mathrm{b}\bar{\mathrm{b}}$
channels were not included yet in the most recent Tevatron combination\cite{TEVhiggs}.

\begin{figure}[htb]
\hfill
\includegraphics[width=.45\textwidth,height=.35\textwidth]{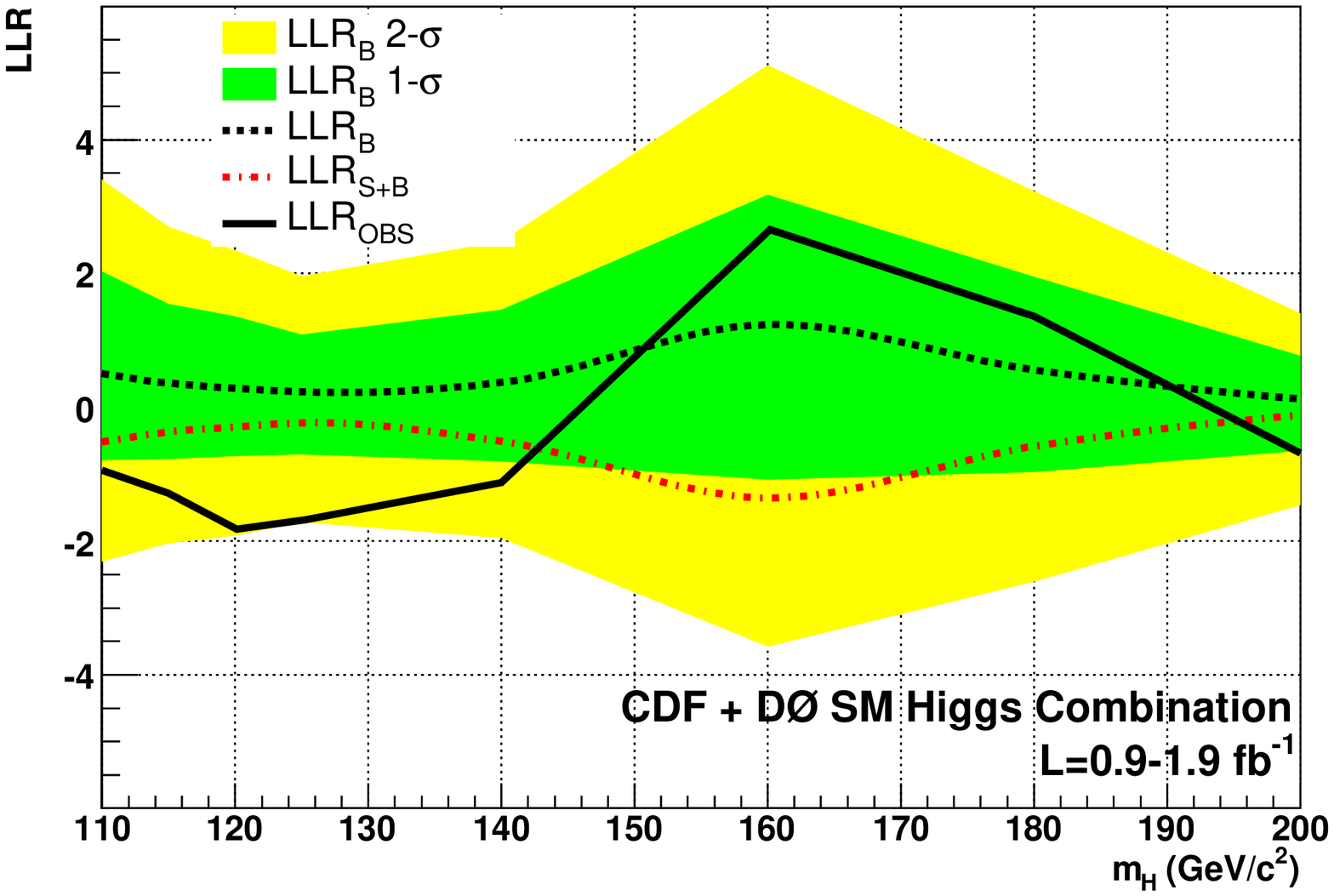}
\hfill
\includegraphics[width=.45\textwidth,height=.35\textwidth]{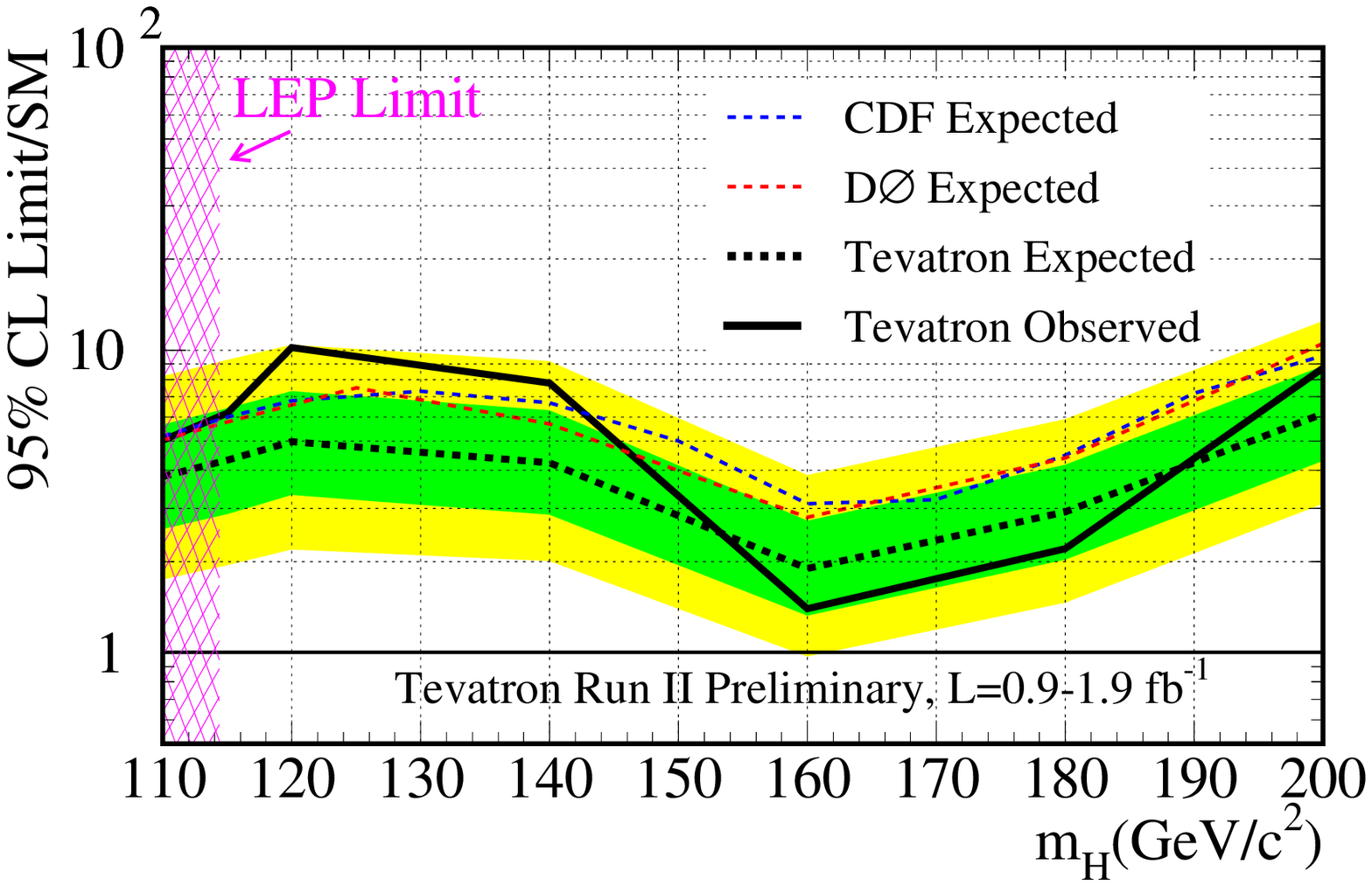}
\hfill
\caption{Tevatron combined Higgs boson search result given as the
  measured and expected log-likelihood-ratio test statistic (left), and
  as the cross section limit divided by the expected standard model
  cross section (right).
\label{fig:limitcombi}}
\end{figure}

The log-likelihood-ratio test statistic 
of the Tevatron combined Higgs boson search is shown
in fig.\ref{fig:limitcombi} (left),
 for both a pure-background
hypothesis (black dashed line) and a signal-plus-background hypothesis
(red dashed line). The separation between these two gives a measure
for the sensitivity, and from the figure it becomes clear that the
sensitivity of the low mass Higgs boson searches is smaller than that for a
high mass Higgs boson. This is mostly due to the different production
and decay processes, and consequently background levels for low and
high mass searches. 
In fig.\ref{fig:limitcombi} (right) the combined cross section limit
is given, as a ratio to the expected standard model cross section. 
At a Higgs boson mass of 115 GeV, the combined limit is a
factor 6.2 (4.3 expected) times the standard model cross section. 

\section{Conclusion and outlook}

In spite of extensive searches the Higgs boson has not been observed
yet by the Tevatron experiments CDF and D0. However, the combined
cross section limit for a Higgs boson mass of 115 GeV is only a factor
6.2 (4.3 expected) away from the cross section predicted by the
standard model. 

The prospects for an exclusion of the existence, or even an
observation of the Higgs boson over a wide mass range are very
good. The data sample accumulated by 2010 is expected to have a size
of 7 to 9 fb$^{-1}$, which is a factor four to eight more than what
was used for the results presented here. Also,
the improvements of the analyses over the last few years have
shown an increase in sensitivity proportional to the accumulated integrated
luminosity. For the near future, this trend is expected to continue,
with additional improvements 
in, e.g., b-tagging, jet energy resolution, multivariate techniques and
lepton identification.

\end{document}